\documentclass[aps,prl,twocolumn,showpacs,superscriptaddress,amssymb]{revtex4}
\usepackage{graphicx}
\usepackage{hyperref}
\usepackage{natbib}
\usepackage{bm}
\usepackage{amssymb} 

\begin{document}

\title{Highly selective detection of individual nuclear spins using the rotary echo on an electron spin as a probe}

\author{V. V. Mkhitaryan}
\affiliation{Ames Laboratory US DOE, Ames, Iowa, 50011, USA}
\author{F. Jelezko}
\affiliation{University of Ulm, Institute of Quantum Optics and Center for Integrated Quantum Science and Technology, 89081 Ulm, Germany}
\author{V. V. Dobrovitski}
\email{slava@ameslab.gov}
\affiliation{Ames Laboratory US DOE, Ames, Iowa, 50011, USA}

\begin{abstract}
We consider an electronic spin, such as a nitrogen-vacancy (NV) center in diamond, weakly coupled to a large number (bath) of nuclear spins, and subjected to the Rabi driving with a periodically alternating phase (multiple rotary echo). We show that 
by switching the driving phase synchronously with the precession of a given nuclear spin, the interaction to this spin is selectively enhanced, while the rest of the bath remains decoupled. The enhancement is of resonant character. The key feature of the suggested scheme is that the width of the resonance is adjustable, and can be greatly decreased by increasing the driving strength. Thus, the resonance can be significantly narrowed, by a factor of 10--100 in comparison with the existing detection methods. Significant improvement in selectivity is explained analytically and confirmed by direct numerical many-spin simulations. The method can be applied to a wide range of solid-state systems.
\end{abstract}
\pacs{76.70.-r,61.05.Qr,76.30.Mi,03.67.Lx}

\maketitle

Detecting a single nuclear spin is an ultimate goal in nuclear
magnetic resonance, actively pursued using various techniques
\cite{mrfm1,wern12,morton,morello,resPRL,Lukin,Zhao12,London}. Moreover, one-by-one detection, characterization, and manipulation of
individual nuclear spins in solids is vital for harnessing them as
a resource for quantum information processing
\cite{gurudev,cappellaro,Slava14,Neumann08,Hanson14,awsch,fuchsPolar,
Maurer12,Laflamme,Cory,smeltzer,ourRev}. The crucial problem, hindering the progress in
this area, is to separate the weak signal, produced by a given weakly coupled nuclear spin, from the much stronger background created by
all other spins and by the ambient magnetic noise. I.e., the detector must possess
high selectivity to separate the signal of the target spin from
the action of other spins and from magnetic noise.
Thus, the focus of the present work is to develop a method for highly selective and sensitive one-by-one detection and manipulation of individual nuclear spins in solids.

It has been recently shown that an electron spin of the
nitrogen-vacancy (NV) centers in diamond is an excellent
nuclear spin detector \cite{resPRL,Lukin,Zhao12,London,DegenImaging,ajoy,kost,Zhao11,
Mamin13,Staudacher13,CJPR2013}.
By applying the dynamical
decoupling protocols to the NV spin (pulse \cite{resPRL,Lukin,Zhao12} or continuous \cite{CJPR2013,London}), the individual
nuclear spins around the NV center can be detected one by one in a
resonant manner: the NV spin becomes decoupled from all nuclear
spins, except the one which satisfies a stringent resonance
condition. The resonating nuclear spin strongly affects the NV
spin motion, and hence can be detected, 
and the width of the
resonances produced by different nuclear spins determines 
selectivity. However, in all current schemes this width is naturally limited, being determined by the
strength of the hyperfine coupling between the target nuclear spin
and the NV spin. 

Overcoming this limitation, and drastically narrowing the resonances in a systematic way, in order to significantly improve the detection selectivity, is of much importance. Here we present a scheme which achieves that goal. It employs the periodically changing Rabi driving on the NV center's spin (multiple rotary echo) \cite{Solomon,Meriles,AielHirCap12} with a specially chosen period to detect the nuclear spins. 
%
The width
of a resonance, corresponding to detection of a given nuclear
spin is inversely proportional to the Rabi driving strength, and therefore can be directly adjusted in experiment. As a result, the resonance peak width is narrowed by a factor of $10-100$ in comparison with the
existing schemes. This narrowing greatly enhances the resolution, which
is very important for accurate characterization of the
nuclear-spin environment of the NV center \cite{Slava14,Neumann08,Zhao12,London,Lukin}, and for
NV-based nuclear magnetic resonance at nanoscale
\cite{Mamin13,Staudacher13,DegenImaging}. Thus, our approach combines excellent protection from magnetic noise offered by the rotary echo, and the greatly enhanced selectivity in detection of individual nuclear spins. This scheme can also be used with other spins as detectors, 
and in some conventional ensemble
magnetic resonance experiments involving electronic and nuclear
spins.

\begin{figure}
\includegraphics[angle=270,width=8cm]{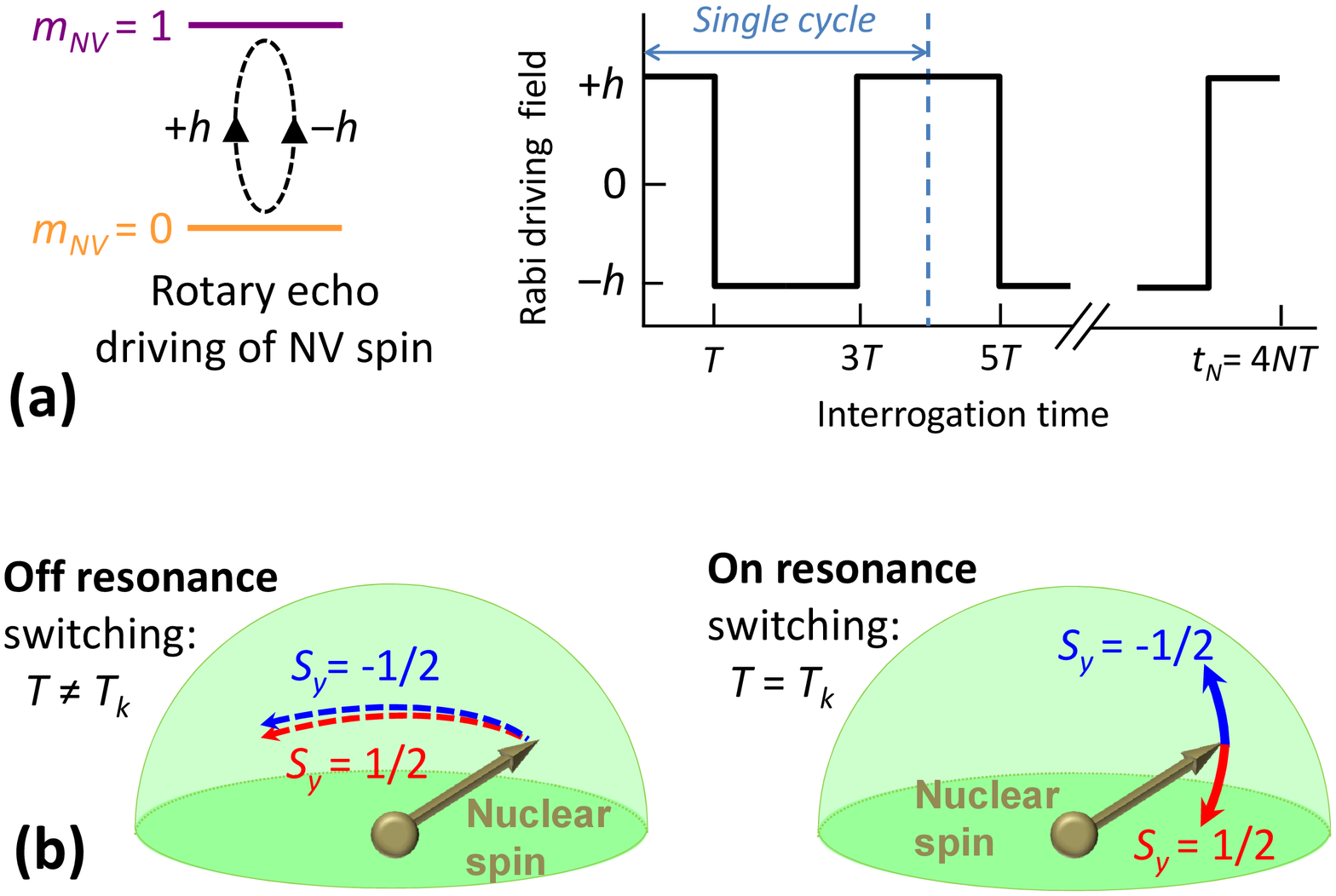}
\includegraphics[width=8cm]{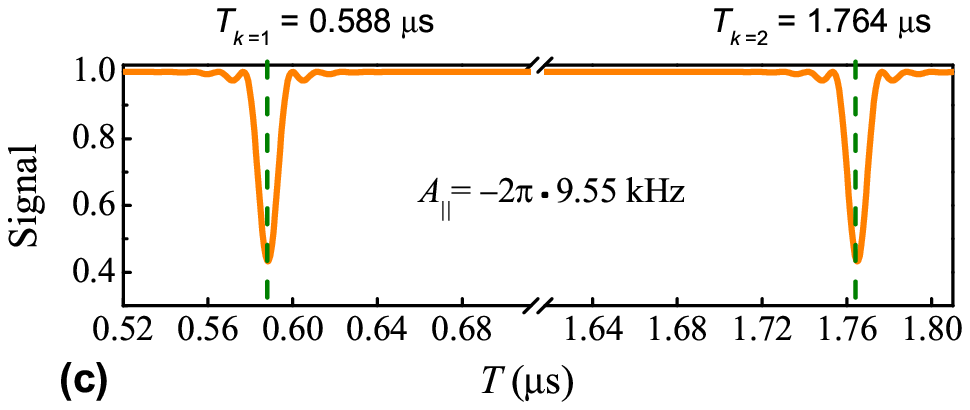}
\caption{\label{fig1}
(Color online) (a) Rotary echo protocol. Rabi driving is applied to the NV electron spin at the frequency of the transition between the states $m_{NV}=0$ and $m_{NV}=1$. The phase of driving  is periodically switched by $180^\circ$, i.e.\ the driving field in the rotating frame periodically changes between $+h$ and $-h$. (b) Generally, the rotary echo protocol efficiently decouples the nuclear spin from the NV center: evolution of the nuclear spin is practically independent of the NV spin state (left). However, when the switching is on resonance with the nuclear spin precession ($T=T_k$), the nuclear spin entangles with the NV spin: its evolution is conditioned on the initial NV spin state (right).
(c) When $T$ becomes equal to the resonance value $T_k$, entanglement with the nuclear spin leads to faster decay of the NV electron spin, seen as a resonant drop of the signal. Graph shows the rotary echo signal (defined everywhere in the paper as $2\langle S_z(N)\rangle$) after $N=50$ cycles as a function of $T$, for
$\omega_L=2\pi\cdot 430$~kHz, driving $h=2\pi\cdot 3.4$~MHz, and coupling constants $A_\parallel=2\pi\cdot -9.55$~kHz, $A_\perp=2\pi\cdot 38.2$~kHz. The resonances of the orders $k=1$ and $k=2$ are marked with green dashed lines.} 
\end{figure}

We consider the electronic spin $S=1$ of the NV center, subjected to a
moderate (tens to hundreds of Gauss) static bias field along the symmetry axis. The NV spin possesses
three well-separated states $m_{NV}=0, 1$ and $-1$ (denoted below as
$|0,\pm 1\rangle$), and is manipulated by applying the
microwave driving at resonance with the transition between the
states $|0\rangle$ and $|1\rangle$ (Fig.~\ref{fig1}a); the level $|-1\rangle$ 
remains idle and will be further ignored. The NV spin is coupled to a nuclear spin $I$: e.g., a $^{13}$C nuclear spins in diamond ($I=1/2$), coupled to the NV center via dipolar interaction. To simplify consideration, we temporarily omit the coupling of the NV spin to the other spins, considering many nuclear spins below (see also \cite{supmat}). In the coordinate frame
rotating with the frequency of the resonant microwave driving, the
system is described by the Hamiltonian
\begin{equation}
\label{eq:NVandI} H = hS_x + \bigl(A_\parallel I_z+ A_\perp
I_x\bigr)S_z + \bigl( \omega_L +A_\parallel/2\bigr) I_z,
\end{equation}
where $S_x=\frac12\bigl(|1\rangle\langle 0|+|0\rangle\langle
1|\bigr)$ and $S_z=\frac12\bigl(|1\rangle\langle 1|-|0\rangle\langle
0|\bigr)$ describe the NV spin, and $I_{x,z}$ are the operators
of the target nuclear spin, $h$ is the Rabi driving field (usually, few MHz to few tens of MHz), $\omega_L$ is the Larmor frequency of the nuclear spin,
$A_\parallel=\omega_h\cos\theta$ and $A_\perp =\omega_h\sin\theta$
are the components of the hyperfine coupling \cite{supmat}, and we assume $\hbar=1$ throughout the text. 

During the rotary echo experiment, the direction of the Rabi driving field (i.e.\ the phase of driving) periodically
changes between $+x$ (along the $x$-axis) and $-x$ (opposite to
the $x$-axis), as shown in Fig.~\ref{fig1}a. Here we consider  symmetrized protocol, 
consisting of $N$ rotary echo cycles; within each
cycle the driving field is along $+x$ during the two outer
segments, each of duration $T$, and is along $-x$ during the inner
segment of duration $2T$. Therefore,
the Hamiltonian (\ref{eq:NVandI}) describes the two outer segments of
the rotary echo cycle, and the Hamiltonian for the inner segment
is obtained by replacing $h\to -h$. Below, we denote these two
Hamiltonians as $H_+$ and $H_-$, respectively, so the evolution
operator for a single cycle is 
\begin{equation}
\label{eo} U=\exp(-iT H_+) \exp(-2iT H_-)\exp(-iT H_+),
\end{equation}
and the full $N$-cycle protocol is described by the evolution
operator $U(N)=U^N$.
Generally, periodic switching of the Rabi field in the regime of
strong driving ($h\gg\omega_h$) leads to highly
efficient decoupling of the NV electronic spin from the
surrounding spins \cite{our,Meriles,AielHirCap12}. However, in a special
resonant regime, where the Rabi field is switched synchronously
with the Larmor precession of the target nuclear spin, quantum
states of the NV center spin and the target nuclear spin become
entangled. After many rotary-echo cycles,
$N\gg1$, this entanglement becomes detectable despite the weak
coupling between the NV and target spins.

To qualitatively understand this phenomenon, we analyze the evolution 
of the system assuming small coupling ($\omega_h\ll h$), so that the zero-order Hamiltonians are 
$H^{(0)}_\pm = \pm hS_x + (\omega_L +A_\parallel/2)I_z$. The corresponding zero-order evolution operator is given by
\begin{equation}
\label{LA} U^{(0)}=\exp \bigl[ -i4T( \omega_L +A_\parallel/2)I_z
\bigr] =\cos2\varphi-2iI_z\sin2\varphi,
\end{equation}
where $\varphi=(\omega_L +A_\parallel/2)T$. I.e., the nucleus and the NV center are practically decoupled: over one cycle, the nuclear spin rotates around the $z$-axis by the angle $2\varphi$, and this rotation does not depend on the state of the NV spin. However, if the driving is switched in resonance with the nuclear spin precession, so that the switching time has a special value, 
\begin{equation}
\label{Tk} T=T_k=\frac{\pi(2k-1)}{2\omega_L+A_\parallel},\quad
k=1,2,...,
\end{equation}
then $\sin2\varphi=0$, and the operator $U^{(0)}$ is unity, i.e.\ the effect of the zero-order Hamiltonian is null after one cycle. Then the smaller higher-order corrections will become important, since their effect will accumulate over many rotary echo cycles, undisturbed by the zero-order terms.
The detailed analysis \cite{supmat} shows that in this resonance case the nuclear spin evolution changes drastically: the rotation axis tilts towards $y$ axis, the rotation angle per cycle is of order of $A_\perp/h$, and most importantly, the nuclear spin entangles with the NV spin through the operator $S_y=\frac{i}{2}\bigl(|0\rangle\langle 1|-|1\rangle\langle 0|\bigr)$. I.e., at resonance the nuclear spin rotates in different directions depending on whether the NV spin is in the state $|+\rangle$ or $|-\rangle$, where $|\pm\rangle$ are the eigenstates of $S_y$ with the eigenstates $s_y=\pm 1/2$, respectively.

We consider an experiment where the NV spin is prepared in the state $m_{NV}=1$, and the operator $S_z(N)$ is measured after $N$ rotary echo cycles. At resonance, the initial coherent superposition 
$|1\rangle=(|+\rangle + |-\rangle)/\sqrt{2}$ is decohered in the basis $|\pm\rangle$ as a result of entanglement with the unpolarized nuclear spin, so that the resonance can be detected by sudden onset of strong decay of $S_z(N)$
%
%
%
%
%
%
Analytical solution, valid for long times ($N\gg 1$), is derived in Supplementary Material \cite{supmat}:
\begin{equation}
\label{eq:resonance} {\rm Signal} = 2\langle S_z(N)\rangle= 1- \frac{2g^2\mu^2}
{\sin^22\varphi + g^2\mu^2 }L(N),
\end{equation}
and in the vicinity of the resonance 
$g\mu\approx(A_\perp/h)\sin{\varphi}\cos{\phi}$ and $L(N)\approx \sin^2{(N\,g\mu)}$, where $\phi=hT$. Thus, the signal as a function of the switching time $T$ (or the angle $\varphi$) has Lorentzian shape with the width $\Delta\varphi=(A_\perp/h)\cos{\phi}$ and the depth 
$2\sin^2{[N(A_\perp/h)\cos{\phi}]}$. To maximize the depth, the driving should be adjusted to have
$\phi=hT=\pi m$ with integer $m$. This also minimizes interference between different nuclear spins and the influence of the fluctuations in the driving power, see Supplemental Information \cite{supmat}, and below we always assume this condition satisfied. 
Typical dependence of the signal on $T$ is shown in Fig.~\ref{fig1}c, where the resonances at $T_{k=1}$ and $T_{k=2}$ are clearly seen. The analytical solution (\ref{eq:resonance}) is very precise: it practically coincides with the direct numerical simulation of the two-spin system described by the Hamiltonian (\ref{eq:NVandI}).

\begin{figure}
\includegraphics[width=8cm]{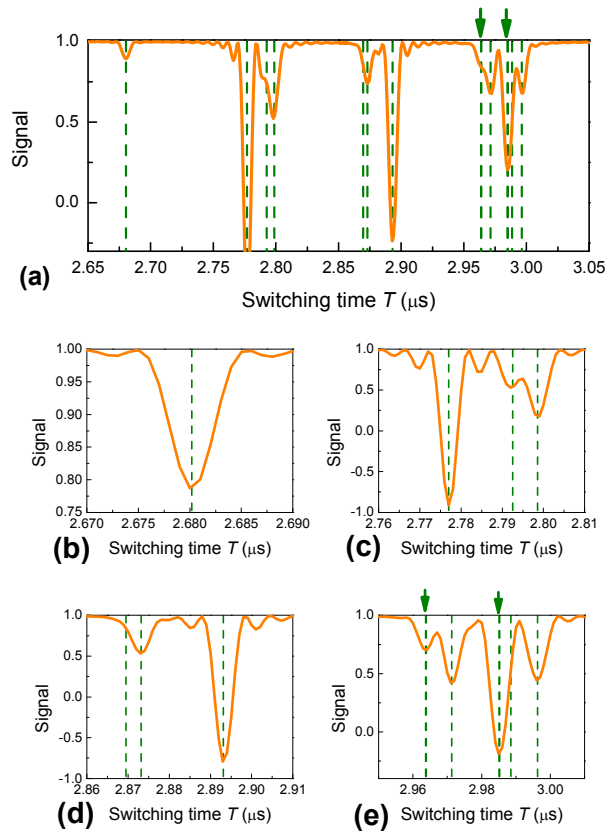}
\caption{\label{fig3}
(Color online) (a) Rotary echo signal after $N=100$ cycles, as a function of the switching time $T$ for a NV spin coupled to 14 
$^{13}$C nuclear spins, randomly placed in the diamond lattice with  natural abundance, for $\omega_L=2\pi\cdot 428$~kHz. Driving $h$ is adjusted so that $hT=28\pi$ (so that $4.59\le (h/2\pi)\le 5.28$~MHz). The region corresponding to the resonances of the order $k=3$ is shown. Theoretically expected resonances are marked with green lines; each of the lines marked with arrows correspond to a pair of nuclei located at the symmetry-related sites, and having almost the same value of $A_\parallel$. Parameters of the nuclear spins are given in Ref.~\onlinecite{supmat}. (b)--(e) Magnified view of the different areas of the panel (a).} 
\end{figure}

For a weakly coupled nucleus, located sufficiently far from the NV center, the hyperfine coupling is determined primarily by dipolar interactions, so that 
$A_\parallel=(\gamma_e\gamma_n/R^3)[1-3R_z^2/R^2]$, where ${\bf R}$ is the vector connecting the positions of the NV center and the nucleus, and $\gamma_e$ and $\gamma_n$ are the electronic and nuclear gyromagnetic ratios, respectively. Therefore, the nuclei from different locations produce the peaks at different values of $T$, and hence can be resolved. 

To test our approach, and to illustrate its performance under realistic circumstances, we performed direct simulations of the rotary-echo detection for the NV center coupled to 14 nuclear spins of $^{13}$C randomly located in diamond lattice. 
The simulation results, given in Fig.~\ref{fig3} for $k=3$, clearly show the sharp, well-resolved peaks, corresponding to different nuclei, with selectivity in sub-kHz region, even for very modest driving $h\sim 2\pi\cdot 5$~MHz (see also Supplemental Materials for more details). Some nuclei, located at the symmetry-related positions in the lattice, have the same $A_\parallel$ and hence the same resonance value of $T$; to resolve them, the static bias field should be tilted away from the symmetry axis of the NV center \cite{Lukin}. The simulations also confirm our conclusion \cite{supmat} that the NV-mediated interaction between the nuclear spins, caused by the driving, does not noticeably affect the detection efficiency.


For a fixed total interrogation time $t_N=4NT$, the resonance width determines selectivity \cite{comment} of the protocol, i.e.\ the ability to resolve two nuclei with close hyperfine couplings $\omega_h^a$ and $\omega_h^b$ (i.e.\ $\omega_h^{a,b}\sim\omega_h$, $|\omega_h^a-\omega_h^b|\ll\omega_h$). High selectivity of the suggested scheme is determined by the small width of the resonances, $\Delta T=A_\perp/(h\omega_L)\sim\omega_h/(h\omega_L)$; here and below we neglect the difference between $A_\parallel$ and $A_\perp$, taking $A_\parallel\sim A_\perp \sim\omega_h$. The resonance width $\Delta T$ should be compared to the distance between the $k$-order resonances from two nuclear spins, which is $\sim (2k-1)(\omega^a_h-\omega^b_h)/\omega_L^2$. Thus, the resonances are resolved when $(\omega^a_h-\omega^b_h)/\omega_h \sim[\omega_L/h]/(2k-1)$. For a typical experiment, detecting $^{13}$C spins, with the bias field of 400~G ($\omega_L=2\pi\cdot 428$~kHz) and driving of $h=2\pi\cdot 10$~MHz, for $k=3$, the condition of resolved resonances is $(\omega^a_h-\omega^b_h)/\omega_h \gtrsim 1/100$. 

The fact that the resonance width can be adjusted by simply changing the driving $h$ is a key feature of the proposed protocol, which leads to great improvement in selectivity. In the currently existing detection schemes \cite{comment}, the width of the resonance is naturally limited, being determined by the coupling constants $A_\parallel$ and $A_\perp$. The scheme proposed here is the first, to our knowledge, where the width of the resonances can be tuned at experimentalist's will; this allows narrowing of the resonances by a factor of 10--100 in comparison with other existing schemes (see \cite{supmat} for details), albeit at the expense of the longer  interrogation time. Also, stronger driving greatly improves the coherence time of the NV spin \cite{Meriles,AielHirCap12,our}. Moreover, using the rotary echo protocol with the optimal choice of driving $h$ significantly reduces the impact of slow fluctuations in the driving strength, which is an important limiting factor for the Hartmann-Hahn double resonance detection \cite{London}.

Above we omitted the NV center's own 
$^{14}$N (or $^{15}$N) nuclear spin: we 
assume that it is 100\% polarized in the state with a given $m_I$, and that the driving is applied at the frequency of the corresponding transition. Without such polarization, the hyperfine levels which are not in resonance with the Rabi driving field are quickly decohered, since the on-site hyperfine coupling ($2.15$~MHz for $^{14}$N, and $3.03$~MHz for $^{15}$N) is comparable to a typical driving strength, so that overall detection quality degrades \cite{supmat}. Fortunately, polarization of the NV's own nuclear spin can be easily achieved in different ways \cite{JelPolar,NeumannQND,resNat,Slava14,fuchsPolar}, and incorporated in the experimental protocol. 

Concluding, we presented a scheme for one-by-one detection of the nuclear spins weakly coupled to an electron spin (e.g.\ the NV center in diamond), using the rotary echo protocol. 
%
%
The key feature of this scheme is the ability to experimentally adjust, and thus greatly narrow, the width of the resonance peaks corresponding to the detected nuclear spins. When compared with the existing approaches, the scheme proposed here 
provides significant improvement in the selectivity of the nuclear spin detection, albeit being more demanding to the quality of experimental setup (timing precision, fast switching of the driving phase, increased interrogation time, etc.). Moreover, the scheme discussed here, and the considerations above, are not specific to the NV center, and can be applied to other 
electron spins, 
or to conventional ensemble double electron-nuclear resonance experiments \cite{supmat}. The main requirements are: (a) well-defined static hyperfine couplings with anisotropy component, (b) nuclear Larmor frequency sufficiently large in comparison to hyperfine couplings, and (c) sufficiently long coherence times of the electron and nuclear spins; a wide range of solid-state systems satisfy these conditions.

We thank R. Hanson and T. H. Taminiau, and M. Raikh for helpful and important discussions.
This work was supported by the Department of Energy --- Basic Energy Sciences under Contract No. DE-AC02-07CH11358.


\begin{references}
\bibitem{comment} 
The protocols discussed in this work are, in fact, two-dimensional, with the signal depending on some adjustable parameter ($T$ for our scheme, $\tau$ for the pulse dynamical decoupling, etc.), and on the total interrogation time. Ref.~\cite{London} proposes to analyze the latter dependence as a way to increase resolution; this option can also be used within the pulse dynamical decoupling detection, and within the protocol presented here. However, the resulting improvement strongly depends on experimental details, and is beyond the scope of the present paper.
\end{references}
\end{document}